\documentclass[aps,prl,superscriptaddress, showpacs, twocolumn]{revtex4-1}
\usepackage{graphicx}
\usepackage{dcolumn}
\usepackage{bm}
\usepackage{amsmath}

\usepackage{color}
\usepackage{bm,graphicx,hyperref}
\hypersetup{%
  breaklinks = {true},
  citecolor = {blue},
  colorlinks = {true},
  linkcolor = {red},
}

\begin{document}

\title{Forward and Inverse Design of Kirigami via Supervised Autoencoder}

\author{Paul~Z.~Hanakata}
\affiliation{Department of Physics, Harvard University, Cambridge, Massachusetts 02138, USA}

\email{paul.hanakata@gmail.com}

\author{Ekin~D.~Cubuk} \affiliation{Google Brain, Mountain View, California 94043, USA}

\author{David~K.~Campbell}
\affiliation{Department of Physics, Boston University, Boston, Massachusetts  02215, USA}

\author{Harold~S.~Park}
\affiliation{Department of Mechanical Engineering, Boston University, Boston, Massachusetts 
02215, USA}

\date{\today}
\begin{abstract}
  Machine learning (ML) methods have recently been used as forward
  solvers to predict the mechanical properties of composite
  materials. Here, we use a supervised-autoencoder (sAE) to perform
  inverse design of graphene kirigami, where predicting the ultimate
  stress or strain under tensile loading is known to be
  difficult due to nonlinear effects arising from the out-of-plane
  buckling. Unlike the standard autoencoder, our sAE is able not only
  to reconstruct cut configurations but also to predict mechanical
  properties of graphene kirigami and classify the kirigami with
  either parallel or orthogonal cuts. By interpolating in the latent
  space of kirigami structures, the sAE is able to generate novel
  designs that mix parallel \emph{and} orthogonal cuts, despite being
  trained independently on parallel or orthogonal cuts.  Our method
  allows us to both identify novel designs and predict, with reasonable
  accuracy, their mechanical properties, which is crucial for
  expanding the search space for materials design.
\end{abstract}

\maketitle
\section{Introduction} Recently, there has been growing interest in
investigating the nonlinear mechanics of perforated thin sheets across
length scales ranging from the macroscale~\cite{dias-sm-48-9087-2017,
  rafsanjani-PRL-118-084301-2017, yang-PRM-11-110601-2018,
  moshe-PRL-122-048001-2019} down to nanoscale
systems~\cite{shyu-NatMat-14-785-2015, blees-Nature-524-204-2015,
  hanakata-Nanoscale-8-458-2016, hanakata-PRL-121-255304-2018}.  The
cuts in a thin sheet---known as kirigami cuts---induce buckling and
other motions (e.g. rotations). These mechanisms result in new
properties, such as enhanced
ductility~\cite{hanakata-Nanoscale-8-458-2016} and
auxeticity~\cite{tang-EML-12-77-2017} that are different from the
pristine (cut-free) counterpart. This simple strategy has led to
programable kirigami actuators which are the building blocks of
soft-robots~\cite{rafsanjani-SR-3-7555-2018,
  dias-sm-48-9087-2017}. While many analytic descriptions have been
developed to understand the changes in mechanical behavior due to the
cuts~\cite{dias-sm-48-9087-2017, rafsanjani-PRL-118-084301-2017,
  moshe-PRL-122-048001-2019}, these analytic approaches are used to
describe systems with repeating and uniform cut patterns or to
optimize a specific target property. An analytical model that can
describe how the mechanical properties of kirigami sheets depend on
the interaction of \emph{different} types of cuts has not been
developed.

For the \emph{inverse design} problem, one ongoing challenge for
kirigami structures is in designing them to achieve specific
properties. Most current ML techniques rely on applying ML to select
top candidates from a \emph {fixed}
library~\cite{seko-PRL-115-205901-2015, hanakata-PRL-121-255304-2018,
  gu-EML-18-19-2018}. The usual approach is to perform ``active
learning'' where the model is trained incrementally with data proposed
by the ML~\cite{seko-PRL-115-205901-2015,
  hanakata-PRL-121-255304-2018}, or by training the model with a
significant amount of data to predict top
candidates~\cite{gu-EML-18-19-2018}.  For both approaches, ML (the
``forward solver'') must be applied to the {\it entire} library. Even
when the computational cost of the ML approach is much lower than the
ground truth data generator (physics-based simulations or experimental
data), in a highly complex system with many degrees of freedom, it is
not practical to use ML to calculate the properties of all candidates
to find the best candidates.

In computer vision problems, \emph{ generative models} have shown to
be successful in generating realistic synthetic
samples~\cite{kingma2013auto}. Unlike supervised learning, the
generative models are trained to capture an underlying data
distribution and to learn important features. For instance,
variational autoencoders have been used to capture the important
information from a high dimensional space of the real representation
(e.g. image) within a lower dimensional space, known as the
\emph{latent space}. The latent vectors capture important features,
for instance smiles in facial images, and thus can be used for
interpolation which is useful for generating new synthetic samples.

In optimizing material properties, we often have some key observable
properties, such as ultimate stress and fracture strain. The goal is
to make the learned hidden (latent) variables correlated to the key
properties, so that we can perform optimization in the \emph{latent
  space}, which has a significantly reduced dimensionality compared to
that of the discrete (original) representation of the structures. This
strategy gives a large advantage over performing optimization in the
original representation space, and has recently been
applied for designing materials with a large design space, such as
drugs, organic molecules, and optical
metamaterials~\cite{kim-npjCompMat-4-67-2018, gomez-ACS-4-268-2018,
  noh-matter-1-1370-2019, ma-AdvMat-31-1901111}.

In this paper we propose a \emph{supervised autoencoder} (sAE) for
inverse structural design. We set up our training such that we can
evaluate the effectiveness of interpolation (generating new designs)
within and outside the training domain. First, we find that the sAE is
able to generate designs consisting of mixed cuts even though the sAE
is trained with kirigami structures with only parallel and orthogonal
cuts, which shows the ability of sAE to perform interpolation in the
latent space. Moreover, in the latent space, the sAE captures
similarities and differences between distinct structures with
different cut types whereas the information about cut types is not
provided during the training. As generalization requires diversity in
the training set, we can leverage the ability of the sAE to
distinguish different structures in the latent space to use it as an
exploration strategy to propose designs that lie outside the training
data.
\begin{figure}
\includegraphics[width=8.6cm]{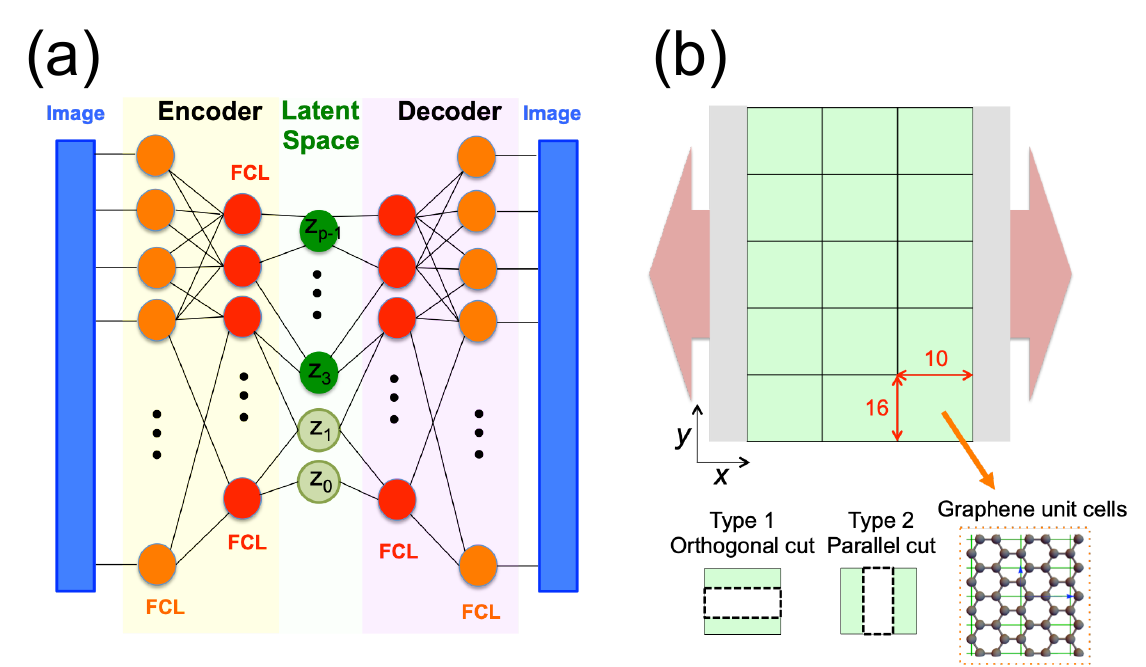}
\caption{(a) Schematic of an autoencoder. (b) Schematic of graphene
  kirigami partitioned into $3\times5$ grids. The training set
  contains either parallel or orthogonal cuts (no mixing of cut
  types). Each grid consists of $10\times16$ graphene unit
  cells. Kirigami is stretched in the $x$-direction by moving the
  edges (grey regions). }
\label{fig:fig1}
\end{figure}

\section{Supervised Autoencoder}An autoencoder (AE) consists of two
parts: (i) an encoder $\mathcal{E}$ that maps a vector to a reduced
representation and (ii) a decoder $\mathcal{D}$ that reconstructs a
vector to its original representation from the reduced
representation. Let $\pmb{x}\in{\rm I\!R^n}=\mathcal{X}$ be the
$n$-dimensional vector, and
$\pmb{z}=(z_0, z_1, \cdots, z_{p-1})\in{\rm I\!R^p}=\mathcal{Z}$ be
the $p$-dimensional latent variables. Since the goal is to have a
compressed representation, $p$ is chosen to be smaller than
$n$. Mathematically we can write this transformation as,
$\mathcal{E}:\mathcal{X}\rightarrow
\mathcal{Z},\,\mathcal{D}:\mathcal{Z}\rightarrow \mathcal{X}$.
In the standard AE the mean reconstruction loss is given by,
\begin{equation}
  \mathcal{L}^{\mathcal{X}}(\pmb{x},\pmb{x}')=\frac{1}{m}\sum_{i=1}^{m}|\pmb{x}^{(i)}-\pmb{x}'^{(i)}|^2=\frac{1}{m}\sum_{i=1}^{m}|\pmb{x}^{(i)}-\mathcal{E}(\mathcal{D}(\pmb{x}^{(i)}))|^2,
\end{equation}
where $i$ is the $i$th data point and $m$ is the number of training
samples. AEs are widely used for unsupervised learning where only
unlabelled data $\pmb{x}$ are provided. In many physical systems, we
want to include the known properties to the unsupervised AE such that
the supervised AE (sAE) learns new ``hidden'' features.  In this work,
we choose $p$ to be 10 and we enforce the first two latent vectors
($z_1$ and $z_0$) to learn ultimate (maximum) stress $\sigma^{\rm u}$
and the corresponding ultimate strain $\epsilon^{\rm u}$. We choose a
latent dimension that is larger than the number of known properties
since kirigami with different cuts can have the same mechanical
properties (e.g. due to symmetries).  Thus, for this proposed
supervised AE (sAE) architecture, we introduce a target property mean
squared error loss function,
\begin{equation}
\mathcal{L}^{\mathcal{Y}}(\pmb{y}, \pmb{z})=\frac{1}{m}\sum_{i=1}^{m}|\pmb{z}^{i}-\pmb{y}^{(i)}|^2=\frac{1}{m}\sum_{i=1}^{m}\sum_{k=0}^{d-1}|z_k^{i}-y_k^{(i)}|^2, 
\end{equation}
where $\pmb{y}\in{\rm I\!R^d}=\mathcal{Y}$ is a $d$-dimensional vector
that contains the known properties and $d$ equals to the number of
observable properties. The total loss function then becomes
$\mathcal{L}=\mathcal{L}^{\mathcal{X}}+\eta\mathcal{L}^{\mathcal{Y}}$,
where $\eta$ is a hyperparameter. We found that $\eta=1$ is a
reasonable choice to get a good accuracy without training the model for
too long. We standardize $\pmb{y}$
to have zero mean and unit standard deviation~\footnote{To normalize
  $\pmb{y}$, we first take the log of $\epsilon^{\rm u}$ and
  $\sigma^{\rm u}$, then subtract from each value its mean, then
  divide it by its standard deviation. We normalize these two
  quantities so that they lie within a similar range.}, which is
essential for training a neural network as the optimizer treats all
directions uniformly in the parameter space~\cite{shanker1996effect,
  mehta2019high}.

In this work, we used the typical AE
architecture~\cite{kramer1991nonlinear}, where a schematic of the sAE
is shown in Fig.~\ref{fig:fig1}(a). For the encoder, we use a deep
neural network (DNN) architecture similar to our previous
work~\cite{hanakata-PRL-121-255304-2018} with one additional
fully-connected (FCL) layer. The decoder consists of two
fully-connected layers. More details of the DNN architecture and
training procedure can be found in the supplemental information (SI).

\emph{Results--}We train the sAE with configurations having parallel
cuts, i.e. that are parallel to the loading direction ($x$-axis), and
orthogonal cuts ($y$-axis), as shown in Fig. \ref{fig:fig1}(b). Each
orthogonal cut has a size of $3\times16$ unit cells (holes), whereas
each parallel cut has a size of $3\times10$ unit cells (holes). We
trained the sAE with configurations having between 0 and 15 cuts.
Each graphene membrane has 2400 unit cells and we define the density
$\rho$ as the number of \emph{holes} divided by the total number of
unit cells. This gives a range of density from 1 (0 cuts) to 0.7 (15
cuts or equivalently 720 holes)~\footnote{The maximum cut density of
  parallel cuts are 15 as there are not detached structures.}. We used
LAMMPS (Large-scale Atomic/Molecular Massively Parallel Simulator) to
simulate graphene kirigami under tension~\cite{plimptonLAMMPS}. The
molecular dynamics (MD) simulation procedure is similar to our
previous work~\cite{hanakata-PRL-121-255304-2018} and the simulation
details can found in the SI. The sAE takes an image of size 2400
($30\times80$) and outputs an image with the same size. While we train
the sAE with configurations having large cuts ($\sim30$ holes in each
grid), in principle, the sAE can generate configurations with any
arbitrary cut size, i.e. as small as one hole. We simulated all
possible configurations of parallel and orthogonal cuts \emph{without}
mixing the two types. As we allow either only 0-15 orthogonal cuts or
0-15 parallel cuts, we obtain a total of 62,558 configurations, of
which 29,791 are non-detached configurations with orthogonal cuts
while the remaining are the configurations with parallel cuts which
have non-detached configurations. The networks were trained with 50\%
of the data set while the remainder of the data set is used for
validation and test set (25\% each).
\begin{figure}
\includegraphics[width=8.6cm]{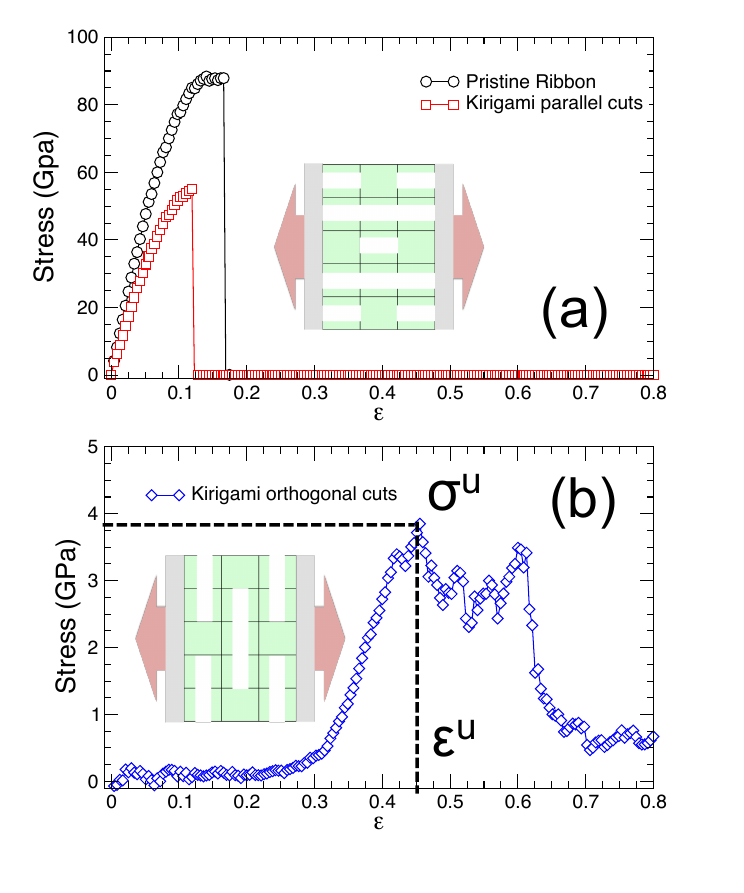}
\caption{Stress as a function of strain for pristine graphene ribbon and
  kirigami with parallel cuts (a) and kirigami with orthogonal cuts
  (b). The stress-strain curve changes significantly as the
  orientation of the cuts is changed. } 
\label{fig:fig2}
\end{figure}

We first show that the mechanical properties of cut graphene are
indeed highly dependent on the material architecture. In some
composite designs, the stiffness of materials can be well-described by
density. In contrast for kirigami, the nonlinear regime becomes
important and thus predicting properties beyond the linear regime,
such as ultimate stress and yield strain, via density is no longer
viable~\cite{hanakata-PRL-121-255304-2018,
  hanakata-Nanoscale-8-458-2016, moshe-PRL-122-048001-2019}. For
instance in typical kirigami geometries, the effective stiffness in
the post-buckling regime is proportional to bending rigidity as
opposed to the Young's modulus~\cite{moshe-PRL-122-048001-2019,
  moshe-PRE-99-013002-2019}.
 
\begin{figure*}
\includegraphics[width=17.2cm]{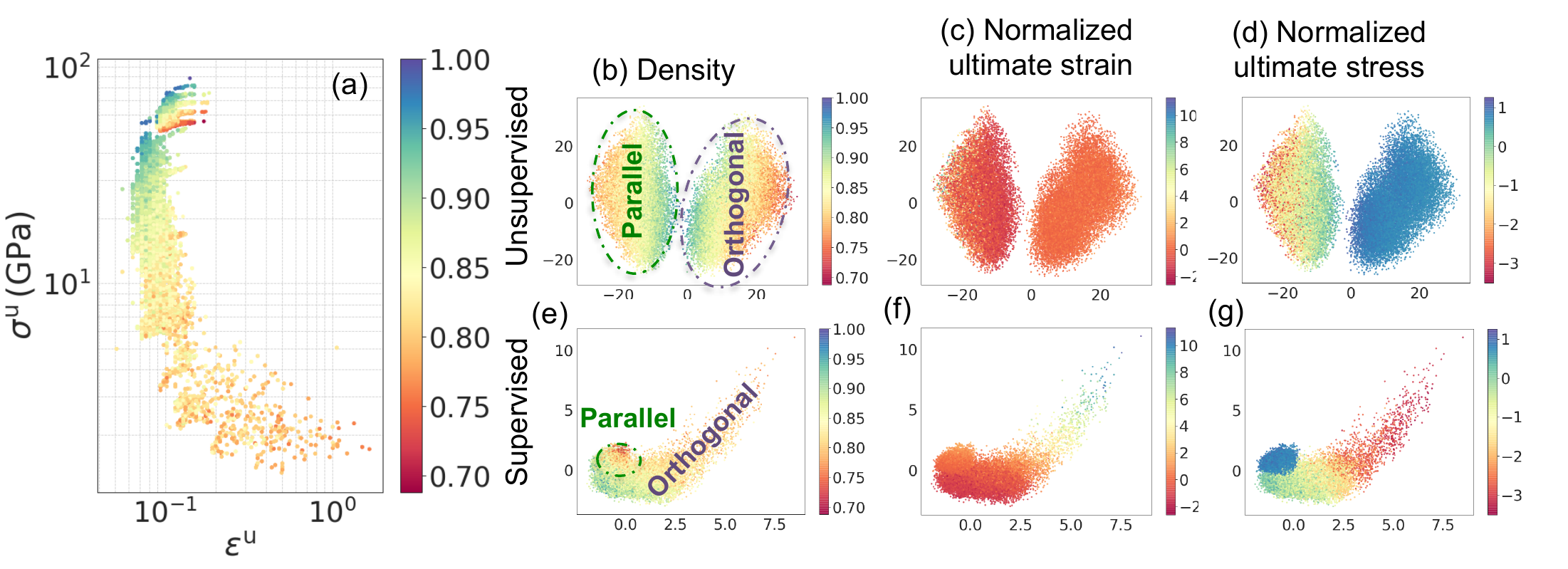}
\caption{ (a) Log-log plot of ultimate stress as a function of
  ultimate strain for \emph{all} simulated data.  The colorbar represents the
  density, where density alone does not correlate with ultimate
  strain. Projected latent space for (b)-(d) unsupervised autoencoder
  and (e)-(g) supervised autoencoder.  The two axis are found by
  PCA. The unsupervised autoencoder is able to distinguish different
  structures based on their cut density and cut orientations. The
  supervised autoencoder successfully captures not only the structural
  differences but also mechanical properties such as normalized
  ultimate strain and normalized ultimate stress. }
\label{fig:fig3}
\end{figure*}

Thus, the \emph{architecture} of the materials strongly impacts their
properties.  To demonstrate this, in Fig.~\ref{fig:fig2}(a) and (b),
we plot stress-strain curves of kirigami with parallel and orthogonal
cuts.  Importantly, the orthogonal cuts in Fig.~\ref{fig:fig2}(b)
represents the same cut pattern with the same number of cuts as shown
for the parallel cut kirigami in Fig.~\ref{fig:fig2}(a). As can be
seen, simply changing the orientation of each cut but fixing the cut
configurations results in a completely different stress-strain curve,
consistent with MD simulations by
Ref.~\cite{zheng-carbon-155-697-2019}.  Furthermore, we can see from
Fig.~\ref{fig:fig3}(a) that density alone does not correlate to
$\sigma^{\rm u}$ or $\epsilon^{\rm u}$.  This further suggests that
the desired global properties are highly dependent on the structural
configuration.

To summarize, the mechanical properties of graphene kirigami depend
not only on (i) material density but also on (ii) cut configurations,
and on (iii) cut orientations. We will show that despite this
complexity our sAE is able to organize the materials based on the
structural properties that are not encoded to the latent space in a
supervised fashion.

Next, we investigate the learned latent variables. We first turn off
the constraint $\mathcal{L}^{Y}$ ($\eta=0$) to enable the AE to learn
in an unsupervised manner. To better visualize the 10D latent space we
project the latent vectors to a 2D space using principal component
analysis (PCA). The latent vectors are generated by passing $\pmb{x}$
of the training data through the encoder. From Fig.~\ref{fig:fig3} we
see that the 2D projected latent variables (from the training data)
clearly separate the two different cut orientations despite the fact
that the AE was not provided with the cut orientations. In addition to
separating structures based on cut orientation, the AE clusters
different structures based on their density. This is similar to how a
latent variable found by AE coincides with the net magnetization (the
order parameter) in the Ising spin
system~\cite{wetzel-PRE-96-022140-2017}. However, none of the latent
variables found by the AE strongly correlates to either
$\sigma^{\rm u}$ or $\epsilon^{\rm u}$ in this kirigami problem.

We now include the property predictions into the latent space.
Similar to the unsupervised AE, as shown in Fig.~\ref{fig:fig3} the
sAE clusters the data based on cut orientation.  Furthermore, by color
coding the data by the normalized ultimate strain
$\overline{\epsilon^{\rm u}}$, and normalized ultimate stress
$\overline{\sigma^{\rm u}}$, we see that in each phase the sAE
organizes structures based on their properties. This shows that the
sAE has not only learned to distinguish different structures of the
input image in the real representation but also to predict their
mechanical properties.  We use the $R^2$ metric to quantify the
performance of the model in predicting $\overline{\epsilon^{\rm u}}$
and $\overline{\sigma^{\rm u}}$ as we did in our previous
work~\cite{hanakata-PRL-121-255304-2018}. The $R^2$ on the training,
validation, test sets for $\sigma^{\rm u}$ ($\epsilon^{\rm u}$) are
0.99 (0.92), 0.99 (0.87) and 0.99 (0.87), respectively. Thus we indeed
find that $z_0$ and $z_1$ are correlated to the normalized
$\epsilon^{\rm u}$ and $\sigma^{\rm u}$, respectively. Similarly, the
sAE is successful in reconstructing the structures $\pmb{x}$ in the
real space. We use the fraction of correctly placed graphene unit cells
as an accuracy metric and we obtain accuracies of 99.4\% for training,
validation, and test set. Details regarding the distribution of all
latent variables and the reconstructed structures of the can be found
in the SI. For the remainder of the paper we will
focus on the sAE.

\section{Generating New Designs via Interpolation in the Latent Space}
While the sAE can be used to generate designs by sampling from the
latent space, the question remains as to how the latent values ($z$s)
are set as they all reside in the same space and are
interconnected. Another simple approach is to perform interpolation in
the latent space. In this section, we introduce metrics to quantify
novel designs and show that we can generate new designs while
simultaneously predicting their mechanical properties with reasonable
accuracy. The question we want to address here is what objective
function should be chosen in order to generate new kirigami designs
that were not in the training data.

\begin{figure*}
\includegraphics[width=17.2cm]{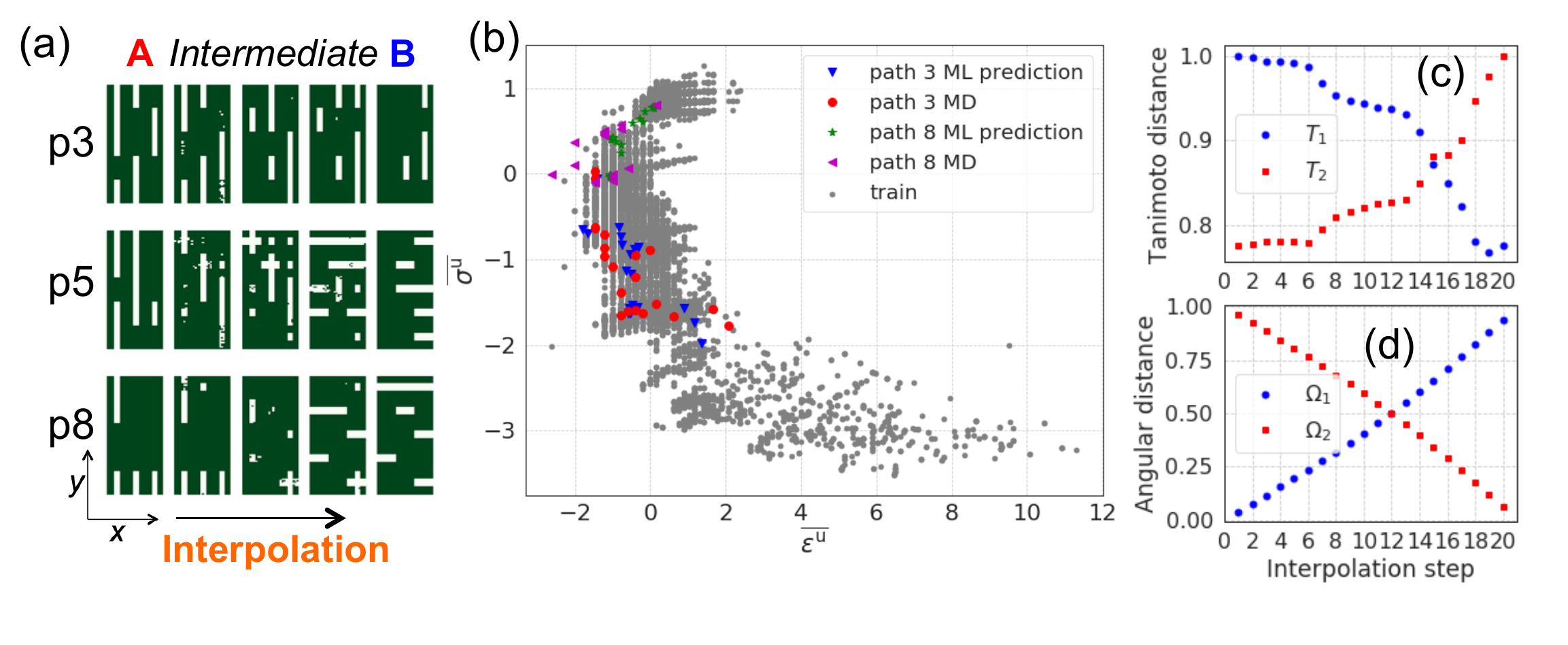}
\caption{(a) Three representative new designs generated by
  interpolating from parent A (first column) to parent B (last
  column). Some structures with mixed cuts are obtained when the two
  parent structures have distinct cut types. (b) Comparison between ML
  predictions and MD results plotted in the normalized ultimate stress
  vs normalized strain plot (mechanical space). Similarity metrics for
  path 8 measured by (c) Tanimoto distance and (d) angular
  distance. Note that the kirigami are stretched in the $x$-direction
  and the edges are not shown. }
\label{fig:fig4}
\end{figure*}

In a $p$-dimensional space, we can write
$\pmb{z}=(z_0, z_1,\cdots, z_{p-1})$ in terms of a radius $r$ and
$p-1$ angles $(\phi_0,\phi_{1},\cdots, \phi_{p-2})$. By analogy to a
genetic algorithm, new designs (children) can be generated by
combining two parents and applying a mutation rule. This approach is
usually performed in the real representation of the genome. In the
current work, we generate new designs from the \emph{latent space},
which is much smaller than the real space. The simplest approach to
generate new designs is by performing linear interpolation between two
latent vectors. Here, we use \emph{spherical linear interpolation}
(SLERP), which has been used for interpolating images in generative
networks~\cite{white-slerp, ha2017neural}. Suppose we have two parent
vectors ${\bf v}_{\alpha}, {\bf v}_{\beta}\in \mathcal{Z}$, then a new
vector can be generated
${\bf
  v}_{t}=\frac{\sin[(1-t)\Omega_{\alpha\beta}]}{\sin\Omega_{\alpha\beta}}{\bf
  v}_{\alpha}+\frac{\sin[t\Omega_{\alpha\beta}]}{\sin\Omega_{\alpha\beta}}{\bf
  v}_{\beta},$
where $0\leq t\leq1$ and
$\Omega_{\alpha\beta}=\cos^{-1}\frac{{\bf v}_{\alpha}\cdot {\bf
    v}_{\beta}}{|{\bf v}_{\alpha}|| {\bf v}_{\beta}|}$.
With this approach the interpolated design lies on the surface of
$p$-dimensional sphere, and the interpolated vector can then be
decoded into a real structure.

As our goal is to perform inverse design outside the training domain,
an important step is to quantify similarity. We use angular distance
($\Omega_{tk}$), and Tanimoto distance ($T_{tk}$) to quantity the
difference between the interpolated structure and the parent
structure:
\begin{align}
\Omega_{tk}&=\cos^{-1}\frac{{\bf v}_t\cdot {\bf v}_k}{|{\bf v}_t|| {\bf v}_k|}/\Omega_{\alpha\beta}\\
T_{tk}&=\frac{{\bf X}_{t}\cdot {\bf X}_{k}}{|{\bf X}_{t}|^2+|{\bf X}_k|^2-{\bf X}_{t}\cdot {\bf X}_{k}},
\end{align}
where $t$ is the interpolation step and $k=\alpha, \beta$. Note that
$\Omega_{tk}\sim0$ indicates two structures that are close in latent
space whereas $T_{tk}\sim1$ indicate structures that are close in real
space.

We generated a total of 200 new structures from 10 pairs of random
configurations obtained from the training dataset. Each interpolation
path contains 20 intermediate structures. We then pass the structures
through the encoder and compare the predicted mechanical properties with
the MD results. The mechanical predictions of half of the 200 new
structures are within 15\% error relative (in real units) to the MD
results. Our discussion will focus on a few representative
examples. Details on how configurations were randomly selected and
results on all other structures can be found in the SI.

Fig.~\ref{fig:fig4}(a) shows intermediate structures from
interpolating two structures with orthogonal cuts (path 3) and
orthogonal and parallel cuts (path 5 and path
8). Fig.~\ref{fig:fig4}(b) shows the corresponding property
predictions and the MD results (path 3 and path 8) in the normalized
ultimate stress vs ultimate strain plot (mechanical space). As shown
in Fig.~\ref{fig:fig4}(a)(p3), the interpolation scheme allows us to
generate similar structures in regions that are close to the training
domain.  It can be seen the MD results are close to the predicted
values. 

In contrast, as shown in Fig.~\ref{fig:fig4}(a) (p5) and (p8), by
interpolating two configurations that have different cut types, we are
able to generate designs consisting of separate parallel and
orthogonal cuts as well as overlapping (mixed) cuts, whereas the
training dataset does not have configurations with two types of
cuts. Because the sAE is interpolating two structures that are
mechanically and structurally different (\emph{far} in the mechanical
space), the predicted mechanical properties are not exact but still in
reasonable agreement. The mean absolute ultimate strain relative error
of the three representative structures are 8.5\%, 51\%, and 15\% for
p3, p5, and p8, respectively. The mean absolute ultimate stress
relative error of the three representative structures are 10\%, 43\%,
and 11\% for p3, p5, and p8, respectively. Several works in computer
vision have also shown that ML models do not generalize well to
samples that are from a slightly different distribution than the
training set~\cite{szegedy2013intriguing, dodge2017study,
  ford2019adversarial}, which means that the ML model can capture only
a subset of the underlying physics. Comparison between MD and ML
predictions for all structures can be found in the SI.

In Figs.~\ref{fig:fig4}(c) and (d) we plot the similarity metrics for
path 8.  We found that designs that are different in real space are
not necessarily different in the latent space~\footnote{One example is
  that two equivalent structures by reflection symmetries (have same
  mechanical properties) will be identified as two different
  structures by Tanimoto distance metric; on the other hand the
  angular distance metric will measure how two designs are different
  structurally and mechanically. }. For instance in path 8, there are
many distinct designs with similar mechanical properties. By comparing
the visualization of the structures and their mechanical properties to
the similarity metrics, we find that the angular distance performs
best in capturing \emph{both} the differences in structures and
mechanical properties.

With this in mind, we recommend using the angular distance as a novel
metric to guide searching in the latent space and to generate new
diverse training data sets or potential designs to obtain
non-redundant models. To show how we can utilize our approach to
search novel designs, we compare two search strategies in generating
structures: (i) select the structure with the highest strain or (ii)
select structure that is the most different, as measured by the
angular distance similarity metric. Out of the 200 representative
generated designs, we obtained 87 designs with mixed cuts when we used
strategy (ii) whereas we only obtained four designs with mixed cuts
when we used strategy (i) (see Figs.~8 and~9 in the SI). It is an open
question in ML research how to best maximize search diversity, which
we hope to further investigate in future studies.

\section{Conclusions}In this work, we have demonstrated the ability of
the supervised autoencoder (sAE) to perform both forward and inverse
design of graphene kirigami.  With regards to forward design, by
distinguishing the difference in mechanical properties depending on
the cut pattern and orientation, the sAE can overcome the traditional
problem of needing to search through the entire design space library
to obtain novel designs.  With regards to inverse design, the sAE
enables the generation of structures by passing the latent variables
to the decoder.  Because the latent space is significantly smaller
than the real space, we can perform optimization in the latent space
as has previously been done to discover new drugs and chemical
compounds~\cite{gomez-ACS-4-268-2018}.  Most importantly, we are able
to classify designs that are different from the training data by
measuring similarity metrics.  While the mechanical property
predictions of the sAE for structures that are significantly different
(far from the training data) are less accurate, the sAE can still be
utilized to propose novel designs.  As online databases for mechanical
systems, such as the mechanical MNIST
database~\cite{lejeune-mechanical-mnist-2020}, are developed, our
model will be important for learning the underlying physics in a
reduced-dimensional space, as well as for proposing novel
designs. Moreover, as the local structures are tightly connected to
electronic properties, this method can be extended for learning
electronic properties in 2D materials, such as pseudomagnetic and
electric polarization, as a function of defects or kirigami cut
patterns ~\cite{zenan-PRB-90-245437-2014, hanakata-PRB-94-035304-2016,
  hanakata-PRB-96-161401-2017, rostami-npj2d-2-15-2018,
  hanakata-PRB-97-235312-2018, torres-PRB-100-205411-2019}.

\begin{acknowledgments}
  P. Z. H. developed the codes and machine learning methods, performed
  the simulations and data analysis, and wrote the manuscript with
  input from all authors. P.Z.H. acknowledges support from the
  National Science Foundation through Grant No. DMR-1608501 and via
  the Harvard Materials Science and Engineering Center through Grant
  No. DMR-1420570. P. Z. H., D. K. C. and H. S. P. acknowledge the
  Boston University High Performance Shared Computing
  Cluster. P. Z. H. is grateful for the Hariri Graduate
  Fellowship. P. Z. H. thank David R. Nelson for helpful discussions.
\end{acknowledgments}
\section{Supplemental Information}
\begin{figure}
\includegraphics[width=8.6cm]{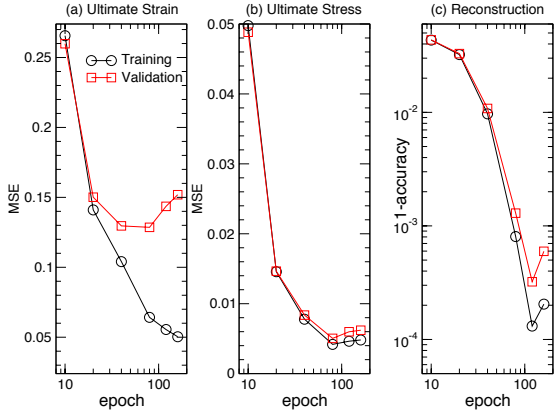}
\caption{Plot of mean squared error (MSE) of (a) normalized ultimate
  strain, (b) normalized ultimate stress, and (c) reconstruction loss
  [$1-{\rm accuracy}$] as a function of epoch number.}
\label{fig:figSI1}
\end{figure}
\subsection{Molecular Dynamics}
We used LAMMPS (Large-scale Atomic/Molecular Massively Parallel
Simulator) to simulate graphene kirigami under
tension~\cite{plimptonLAMMPS}.  The carbon-carbon interactions are
described by the AIREBO potential~\cite{stuart2000reactive}, where the
simulation procedure is similar to our previous
work~\cite{hanakata-PRL-121-255304-2018}.  The lattice constants of
the rectangular graphene unit cell are $\sim4.2$~\AA~ and
2.4~\AA~along the $x$ and $y$ directions, respectively. The graphene
is first relaxed using conjugate gradient energy minimization and the
relaxed within an NVT (fixed number of particles $N$, volume $V$ and
temperature $T$) ensemble for 50~ps with non-periodic boundary
conditions in all directions at a fixed temperature $T=4.2$~K. The
graphene is then stretched by moving the right and left edges at a
constant strain rate of $\sim0.005/{\rm ps}$ which is slower than our
previous work. In this work, we use a
${\rm rcmin}\_{\rm CC}=1.92$~\AA~cutoff for the REBO potential and a
6.8~\AA~cutoff for the Lennard-Jones term in the AIREBO
potential. Other parameter values in the AIREBO potential are kept the
same following the default CH.airebo. The
${\rm rcmin}\_{\rm CC}=1.92$~\AA~cutoff is chosen to avoid artificial
strain hardening behavior~\cite{wei2012nature,
  hanakata-PRL-123-069901-2019}. We use a NVT time damping constant
$t_{\rm damp}=10$. To calculate the 3D stress we multiply the stress
tensor $\sigma_{xx}$ (along the loading) by the virial thickness
(simulation box) and divide it by an effective graphene thickness of
3.35~\AA. Similar procedures have been used for other MD and density
functional theory simulations and
experiments~\cite{zhang-NatComm-5-3782-2014,
  hanakata-Nanoscale-8-458-2016, hanakata-PRB-94-035304-2016,
  lee-science-321-385-2008}.

\subsection{Machine Learning}
We used TensorFlow (version r1.12) to build the sAE~\cite{tensorflow2015-whitepaper}.  We used
scikit-learn~\cite{scikit-learn} for the principle component analysis
(PCA). The TensorFlow r1.12 was run on four CPUs and one NVDIA Tesla
K40m GPU card.

For the encoder, we use a deep neural network (DNN) architecture
similar to our previous work~\cite{hanakata-PRL-121-255304-2018}. The
first part of the DNN consists of three convolutional layers with 16,
32, and 64 filters followed by two fully-connected layers (FCL) of
size 512 neurons and 128 neurons. A stride of 1 and a kernel of size
$3\times3$ are used in the convolutional layers.  For the decoder, we
use two FCLs with 512 and 1024 neurons without convolutional layers. A
schematic of the sAE is shown in Fig.~1(a) of the main text. Each
convolutional layer is followed by rectified linear unit (ReLU)
activation function and max pooling. Each FCL is followed by ReLU
activation functions except for the last layer (output layer) of the
decoder where we use sigmoid activation functions to generate the
reconstructed image (of size $30\times80$), which is a string of 0s
and 1s.  This proposed ML network enables us to simultaneously learn
to predict the desired multi-target properties while also
reconstructing the structures.
\begin{figure}
\includegraphics[width=8.6cm]{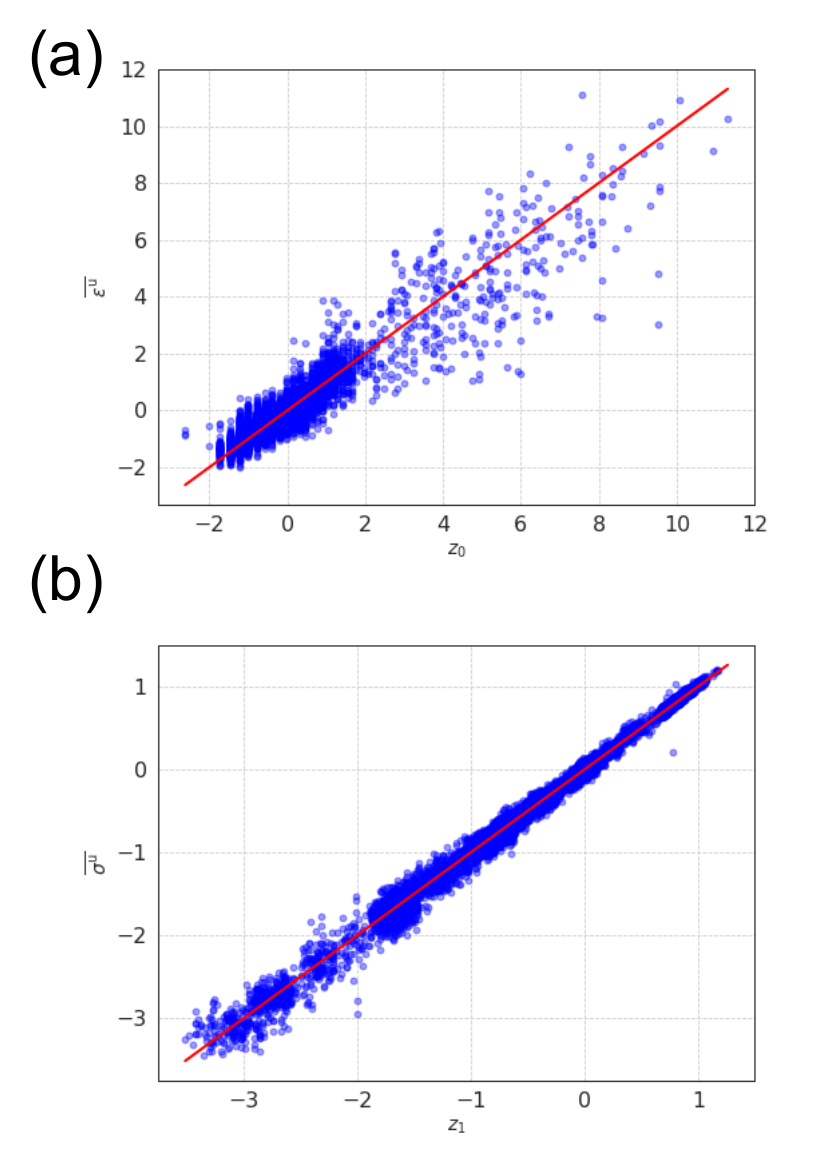}
\caption{Plot of true values as a function of predicted values for (a)
  normalized ultimate strain and (b) normalized ultimate stress (b).}
\label{fig:figSI2}
\end{figure}

Both losses $\mathcal{L}^{\mathcal X}$ and $\mathcal{L}^{\mathcal Y}$
are monitored and the training is stopped when the losses are within
the range of desired accuracy. We used the mean square error loss to
monitor the loss for the ultimate strain and ultimate stress and used
the accuracy score to monitor the performance of the model for
reconstruction. The accuracy score is the ratio of the correctly
placed elements (holes or no holes) divided by the total number of
unit graphene cells. Specifically, we stop the training when the
$\mathcal{L}^{\mathcal{Y}}$ validation loss of $\sigma^{\rm u}$ and
$\epsilon^{ \rm u}$ start increasing and becoming much larger than the
training loss--- an indication of overfitting---and we ensure that the
reconstruction loss $\mathcal{L}^{\mathcal{X}}$ is around 0.001---a
tolerance of misplacing three unit cells out of 2400 unit cells. We
found that training the sAE for too long (too large a number of
epochs) will result in a model with a high reconstruction accuracy but
a poor performance for predicting mechanical properties. We thus
decided that epoch of 80 represented an appropriate balance between
reconstruction and property prediction loss as shown in
Fig.~\ref{fig:figSI1}. In this plot we use sAE with two convolutional
layers followed by two fully connected layers of size 512 and 128
(encoder), a latent dimension of size 10, and two fully connected
layers of size 512 and 1028 (decoder). In addition, we report $R^2$
score to attest the performance of property prediction as we did in
our previous work~\cite{hanakata-PRL-121-255304-2018}. A learning rate
of 0.0001 with a batch size of 128 and epoch 80 were used. The Adam
optimizer was used for minimizing the total loss.

\begin{figure}
\includegraphics[width=8.6cm]{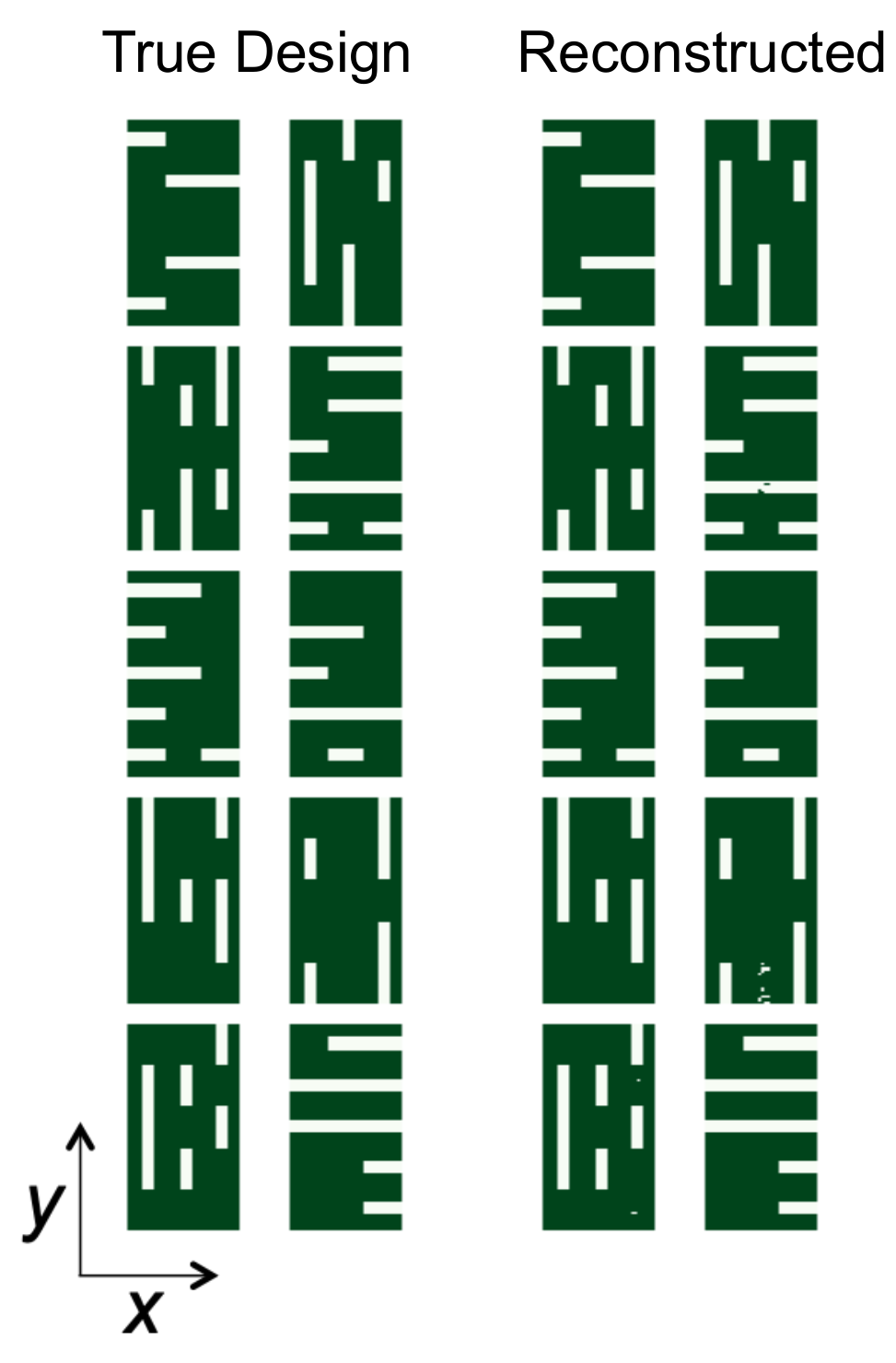}
\caption{Reconstructions of inputs obtained from the test set. The
  sAE can reconstruct the input data closely to the ground truth.}
\label{fig:figSI3}
\end{figure}

In Fig.~\ref{fig:figSI2}, we plot the property predictions of the
supervised autoencoder (sAE) on the \emph{test} set. The $R^2$ on the
training, validation, and test sets for $\sigma^{\rm u}$
($\epsilon^{\rm u}$) are 0.99 (0.92), 0.99 (0.87) and 0.99 (0.87),
respectively. We see that the sAE can reconstruct the test data
(configurations in real space) very close to the ground truth with
some imperfections, shown in Fig.~\ref{fig:figSI3}.  We use the
fraction of correctly placed graphene unit cell as an accuracy metric
and we obtain accuracies of 99.4\% for training, validation, and test
sets. Note that while the ML has never seen the test set, the test set
is taken from a data set that contains either parallel or orthogonal
cuts only. As discussed in the main text, we investigate how well the
ML can generalize for structures \emph{far} beyond the training
samples---such as \emph{mixed} cuts.

\subsection{Latent variable distribution}
\begin{figure}
\includegraphics[width=8cm]{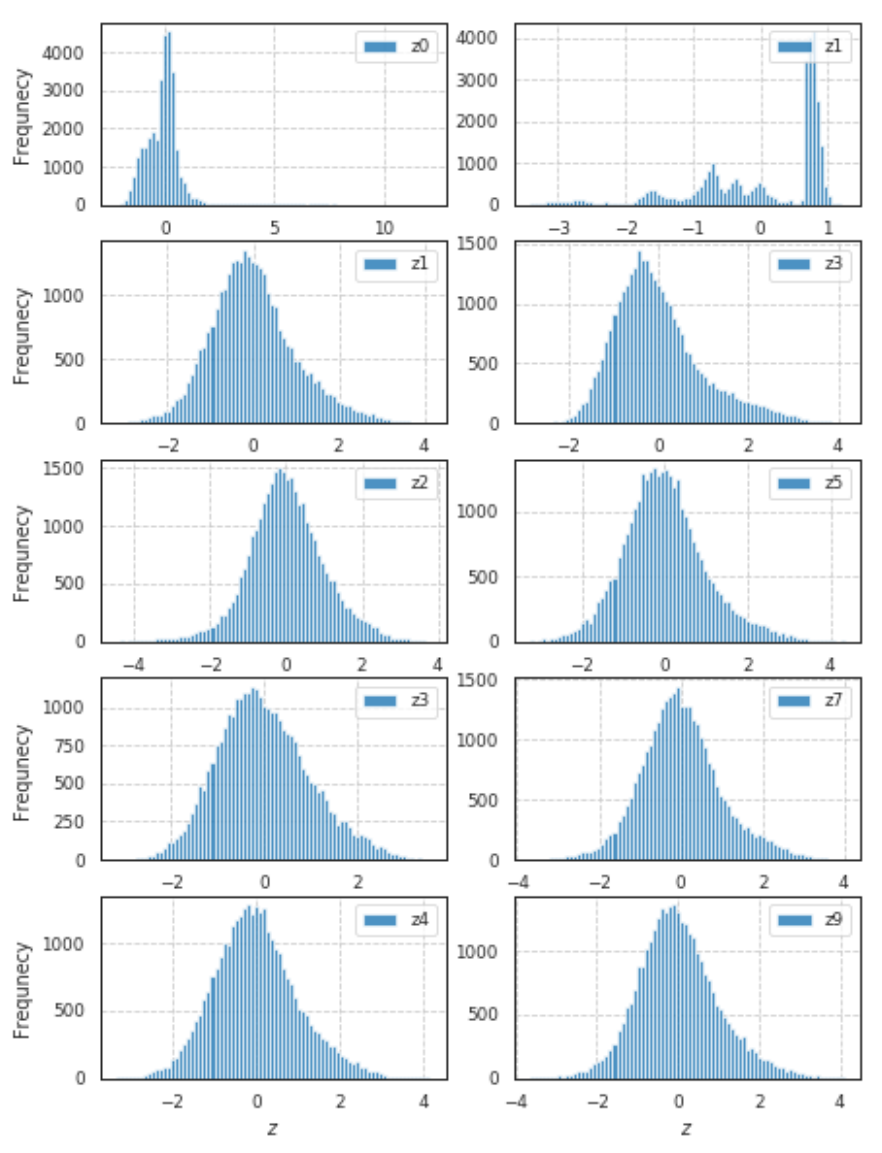}
\caption{Histogram of all latent variables ($z$s) obtained by passing the training data
  through the encoder. }
\label{fig:figSI4}
\end{figure}
\begin{figure}
\includegraphics[width=8cm]{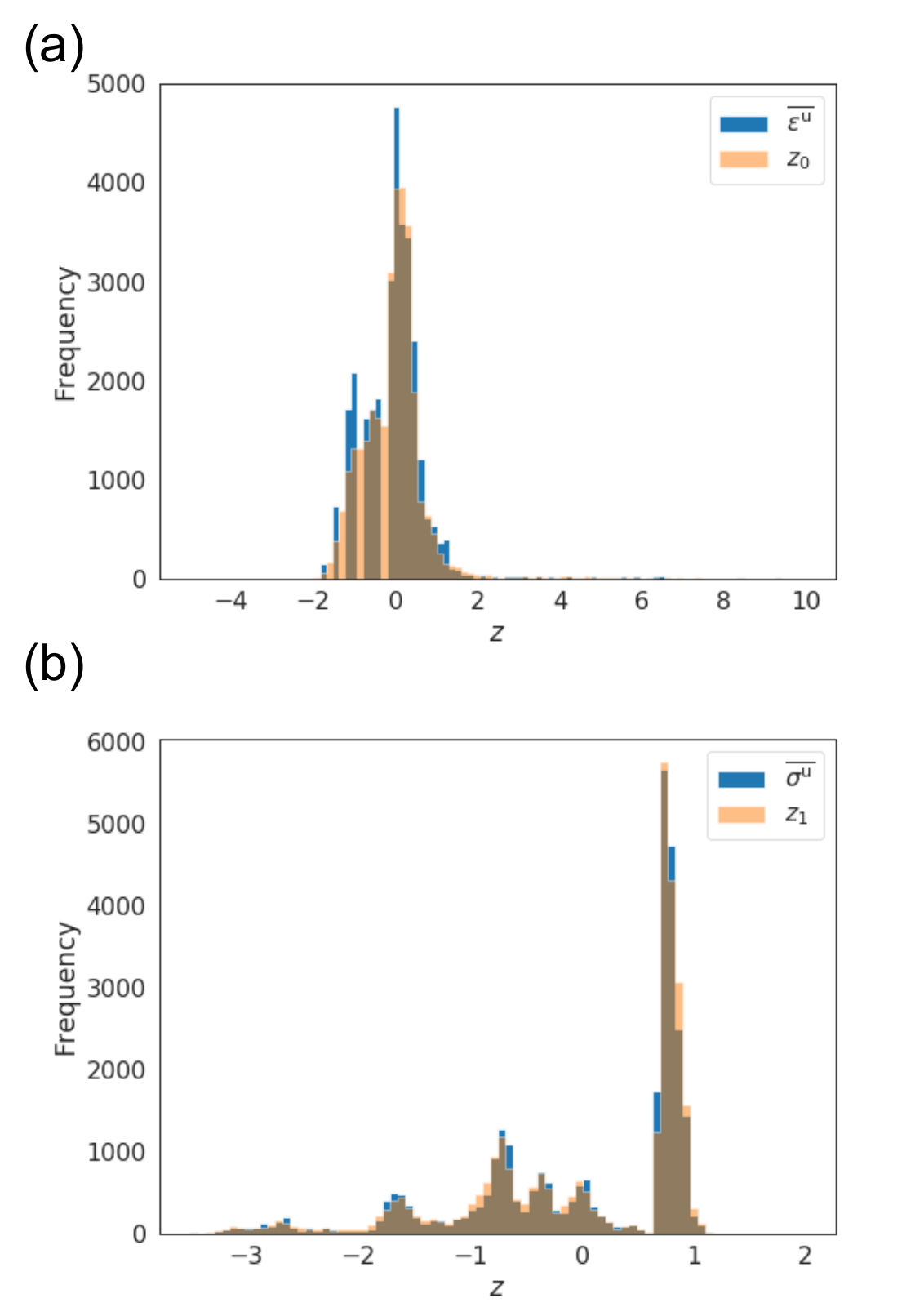}
\caption{Histograms of the true values of normalized yield strain
  $\overline{\epsilon^y}$ and hidden variable $z_0$. (b) Histogram of
  the true values of normalized yield stress $\overline{\sigma^y}$ and
  hidden variable $z_1$. The encoder is trained to predict
  $\epsilon^y$ and $\sigma^y$, and indeed finds similar
  distributions.}
\label{fig:figSI5}
\end{figure}

To find the distribution of latent vectors $z$ we pass the training
set through the encoder to obtain the latent vector for each training
data point (see Fig.~\ref{fig:figSI4}). We see that $z$ can have a
Gaussian distribution.  We find that the sAE can generate non-Gaussian
distributions of $\epsilon^{\rm u}$ and $\sigma^{\rm u}$.
Fig.~\ref{fig:figSI5} shows how the histograms of $z_0$ and $z_1$
closely represent the actual values of $\epsilon^{\rm u}$ and
$\sigma^{\rm u}$.

\subsection{Validating SLERP predictions}
\begin{figure*}
\includegraphics[width=15cm]{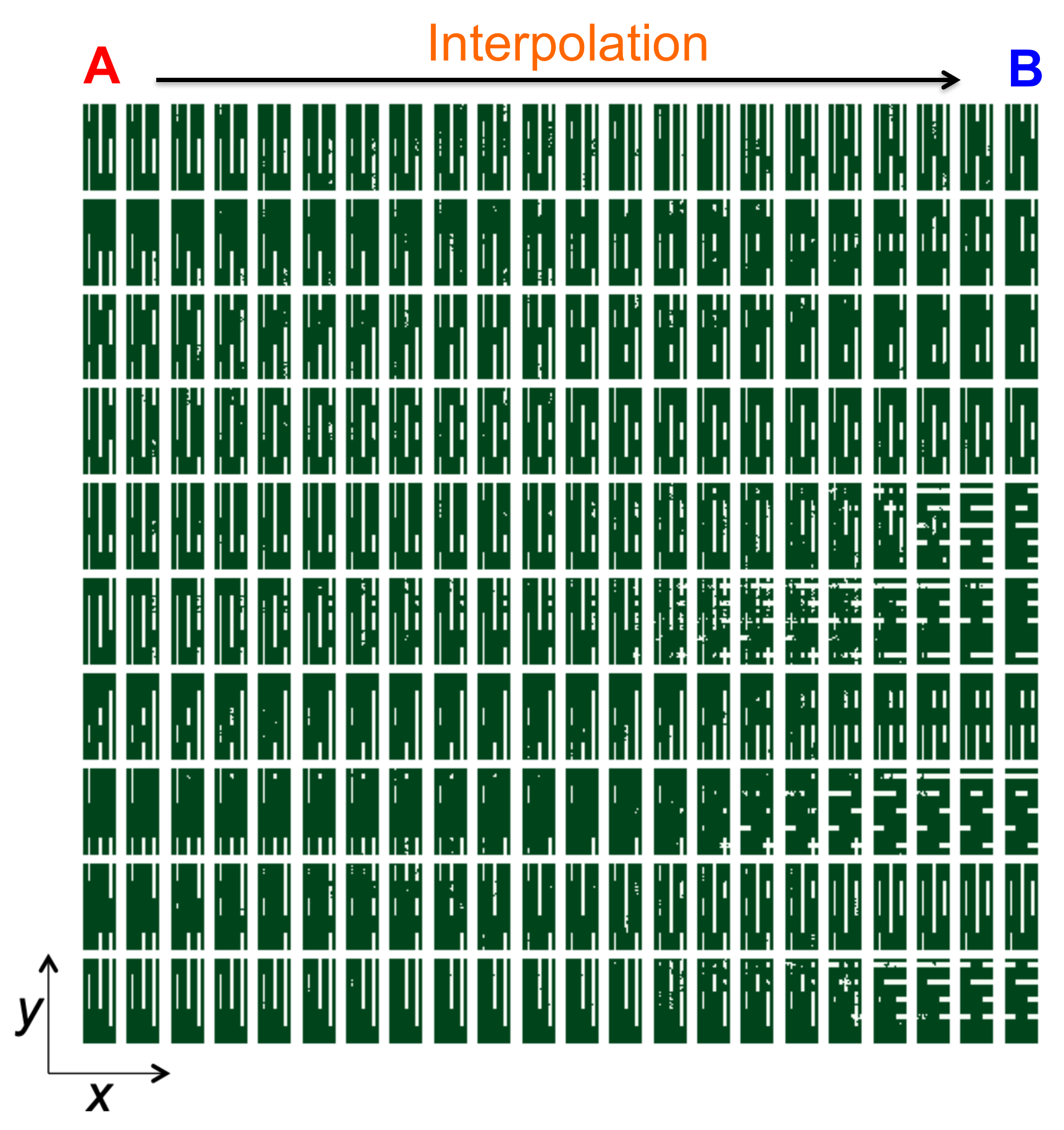}
\caption{New designs generated by interpolating from parent A to
  parent B with 20 steps. Some structures with mixed cuts are obtained
  when the two parent structures have distinct cut types.  Kirigamis
  are stretched along the $x$-axis.}
\label{fig:figSI6}
\end{figure*}
In a $p$-dimensional space, we can write
$\pmb{z}=(z_0, z_1,\cdots, z_{p-1})$ in terms of a radius $r$ and
$p-1$ angles $(\phi_0,\phi_{1},\cdots, \phi_{p-2})$.
\begin{align}
z_0&=r\prod_{k=0}^{p-2}\cos\phi_k\\
z_i&=r\sin\phi_{i-1}\prod_{k=i}^{p-2}\cos\phi_k,\quad i=1,\cdots, p-2\\
z_{p-1}&=r\sin\phi_{p-2} 
\end{align}
Here we compare the MD simulation results and the the prescribed
values (via interpolation). We randomly selected structures from
different domains. Fig.~\ref{fig:figSI6} shows the
structures generated by interpolation in the latent space. In the
first column, the top (bottom) row is the first (second) parent which
is randomly selected from the top 1000 designs sorted based on
$\epsilon^{\rm u}$. In the second column parents are selected from the
bottom 1000. New designs can be generated smoothly by interpolating
two structures that are close in the latent space---i.e. that have
similar $\sigma^{\rm u}$ and $\epsilon^{\rm u}$. As shown in
Fig.~\ref{fig:figSI6}, the interpolation scheme allows us to generate
similar structures in regions that are close to the training
domain. 

Next, in the 3rd to 6th rows, for the first column (first parent)
configurations are randomly selected from the top 1000 of the training
set sorted based on $\epsilon^{\rm u}$, while in the 7th to 10th rows
the first columns configurations are randomly selected from the bottom
1000 sorted based on $\epsilon^{\rm u}$. For the last column (second
parent) in the 3rd to 10th rows, configurations are randomly selected
from the entire training set. We use this selection scheme to show how
the interpolation scheme performs in different regions.
\begin{figure*}
\includegraphics[width=15cm]{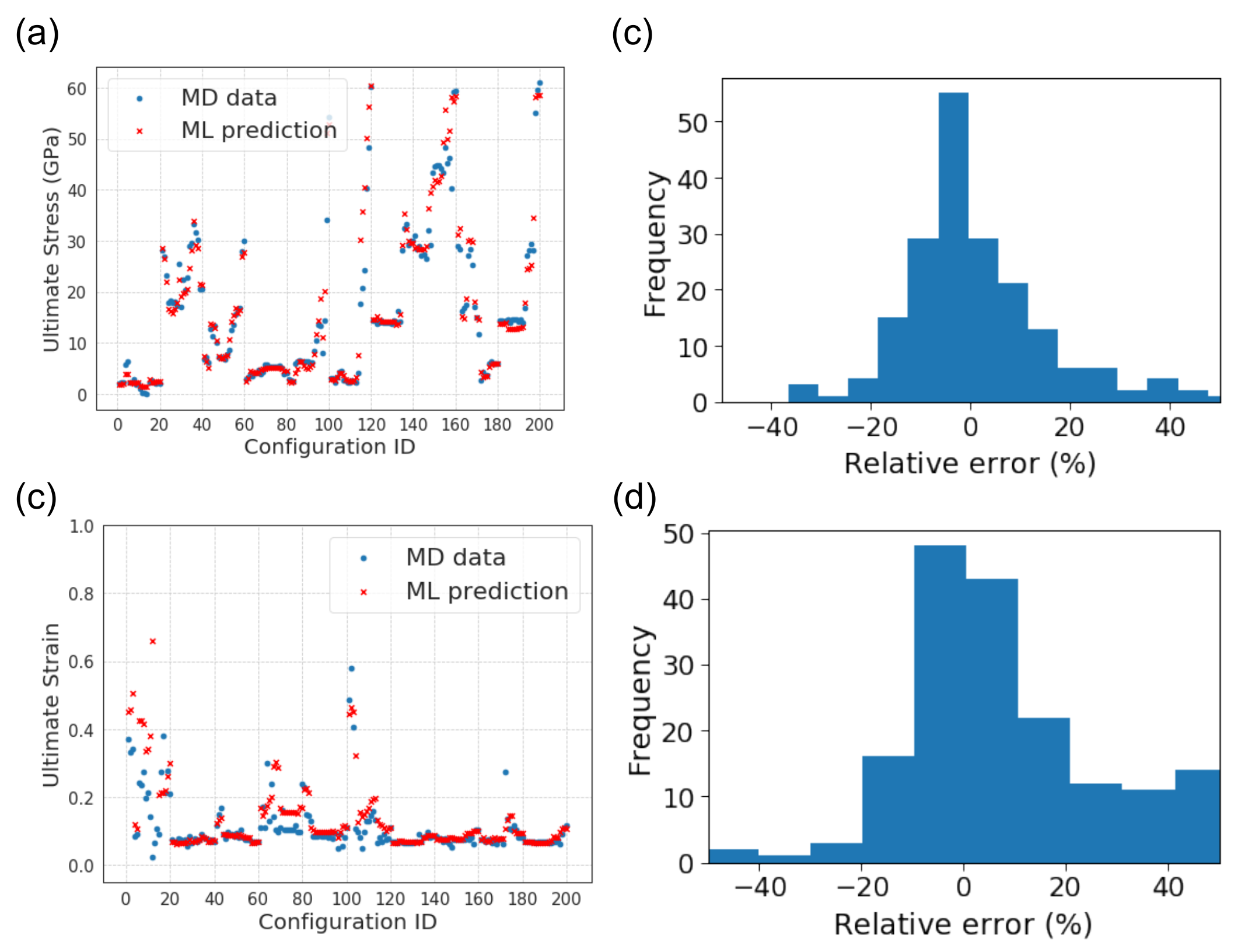}
\caption{Comparison between (a) ultimate strain and (b) ultimate
  stress values of MD results and ML predictions. Configuration ID 1
  to 20 are structures with interpolation path 1 and so on. There are
  10 paths and each path has 20 interpolation steps; this gives a
  total of 200 configurations. Histograms of relative errors for (c)
  ultimate stress and (d) ultimate strain.The corresponding
  interpolated structures are plotted in Fig.~\ref{fig:figSI6}. }
\label{fig:figSI7}
\end{figure*}

Next, we compare the ML predictions and the MD results (see
Fig.~\ref{fig:figSI7} (a) and (b)). For clarity we also plot the
histograms of relative error. Relative error is defined as
\begin{equation}
\text{\rm Relative error}=\frac{y_{\rm ML}-y_{\rm MD}}{y_{\rm MD}}\times100\%,
\end{equation} 
where $y_{\rm ML}$ is the mechanical property predicted by ML and
$y_{\rm MD}$ is the mechanical property obtained from MD
simulations. Note that here we use the real values to compare the
relative error. Fig.~\ref{fig:figSI7}(c) and (d) show the histogram of
relative error of ultimate strain and ultimate stress in real
units. We see that the property predictions of roughly 100 structures
are within 15\% error. Generally, the deviation is highest when the
interpolated structures are at the equidistant point from both
parents. We can utilize this to generate new novel designs where the
ML model are yet to learn a new domain.

\subsection{Generating Diverse Designs}
\begin{figure*}
\includegraphics[width=15cm]{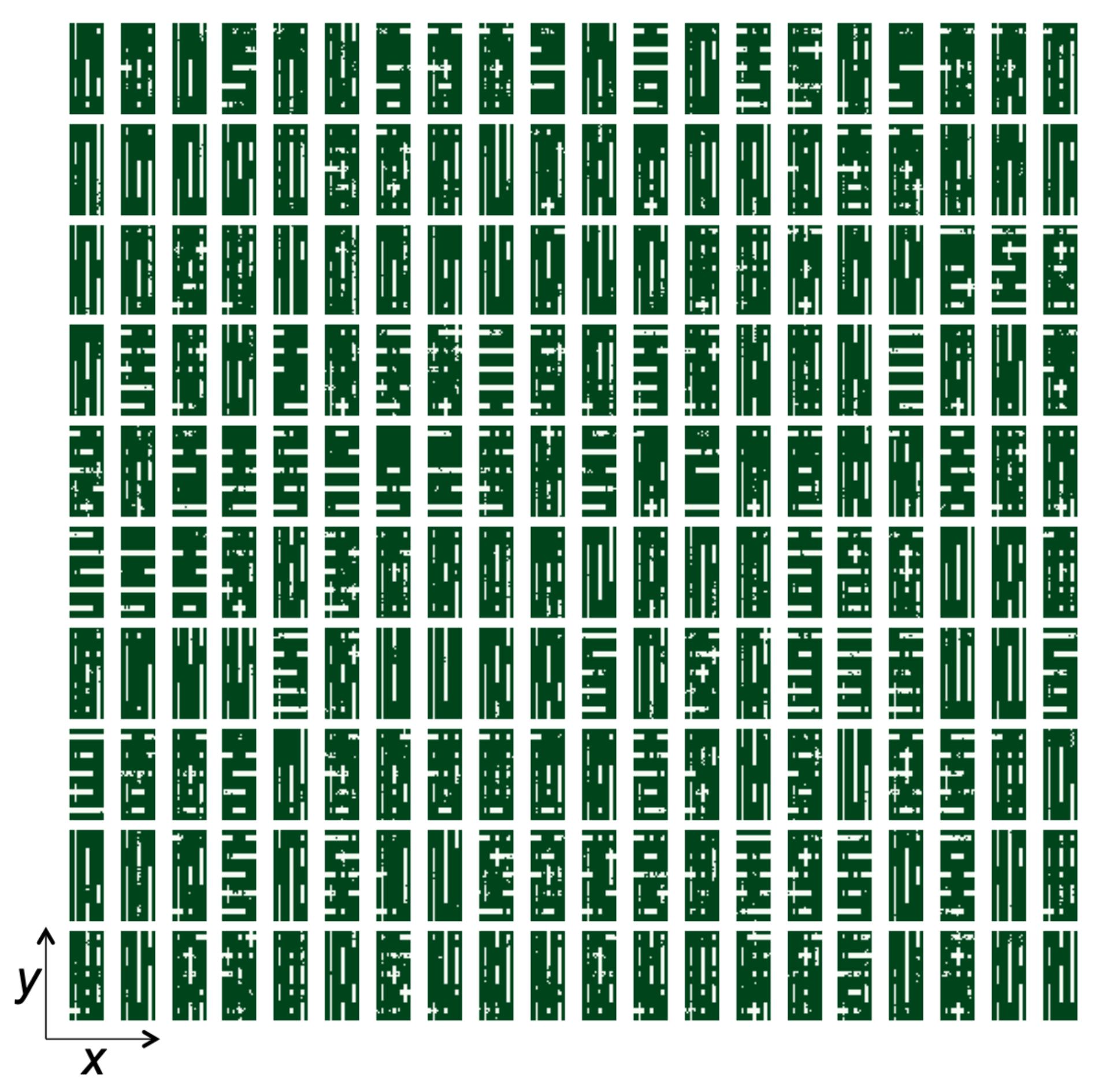}
\caption{Generated structures by maximizing diversity. Kirigamis are
  stretched along the $x$-axis. The relative errors are plotted in
  Fig.~\ref{fig:figSI10} (a) and (b).}
\label{fig:figSI8}
\end{figure*}

\begin{figure*}
\includegraphics[width=15cm]{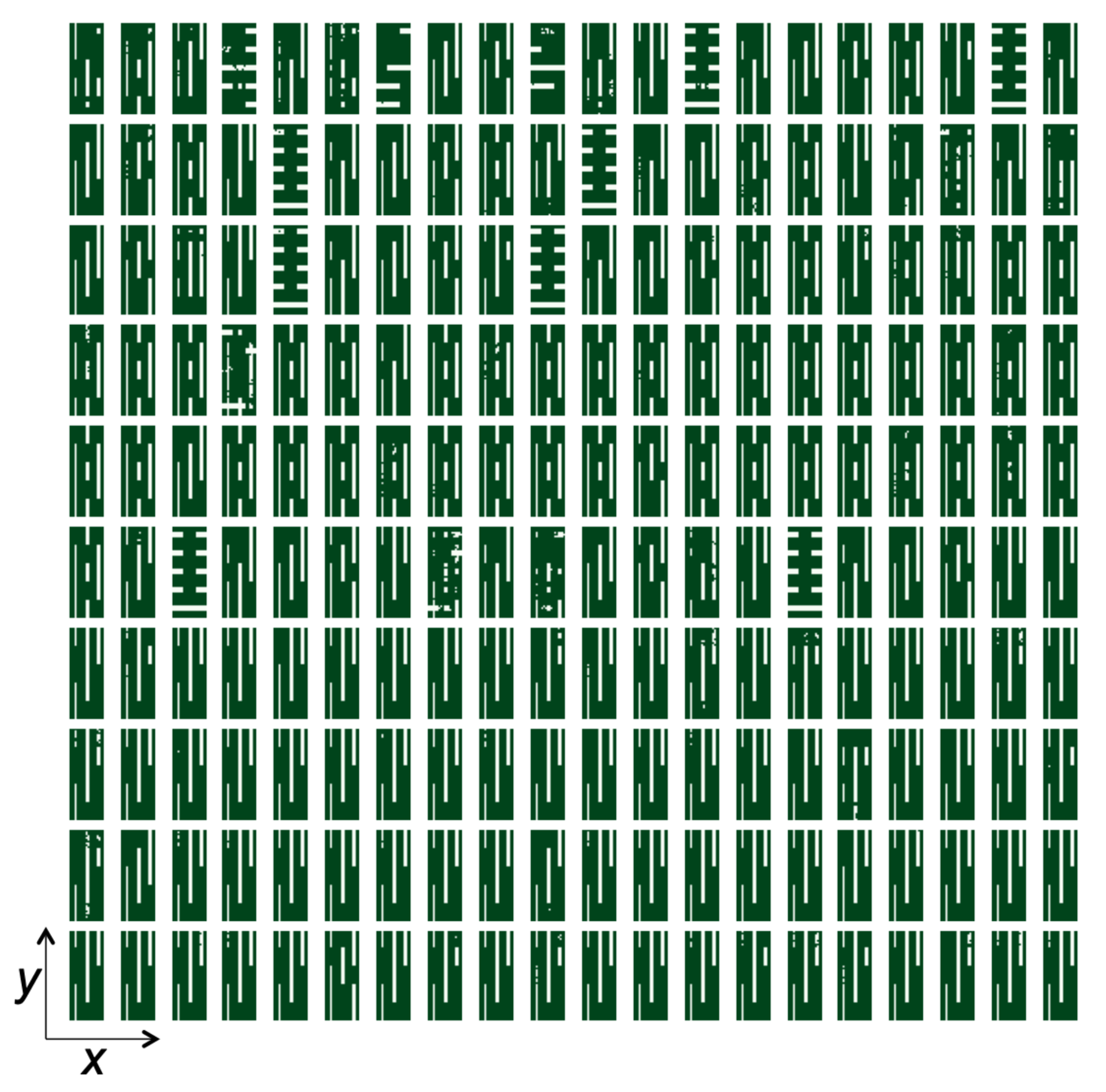}
\caption{Generated structures by maximizing strain.  Kirigamis are
  stretched along the $x$-axis. The relative errors are plotted in
  Fig.~\ref{fig:figSI10}(c) and (d).}
\label{fig:figSI9}
\end{figure*}

In this section we show how we can generate more diverse designs by
choosing interpolated structures that are mostly different from their
parents, as measured by the angular distance. We selected 100 random
configurations from the training set and from this set we obtained
4950 paired parents. We then generated new designs using these 4950
paired parents with 20 interpolation steps. Here we contrast two
search strategies. In the first strategy we only select designs with
highest strain and in the second strategy we select designs that are
the most different from their parents, as measured by the angular
distance. For easier comparison, we selected 200 representative
configurations (out of 4950) with $\overline{\epsilon^{\rm u}}>1$ and
$\overline{\sigma^{\rm u}}>-3$. In Fig.~\ref{fig:figSI8}, we can see
that there are many structures with mixed and overlapping cuts where
the strategy to achieve maximum structural difference was used. On the
other hand using the first strategy (maximizing strain only), we
obtained only two structures containing mixed cuts, as shown in
Fig.~\ref{fig:figSI9}. To conclude, we can use our ML model and the
angular distance metric to generate structures that are different from
the training set.

\begin{figure*}
\includegraphics[width=15cm]{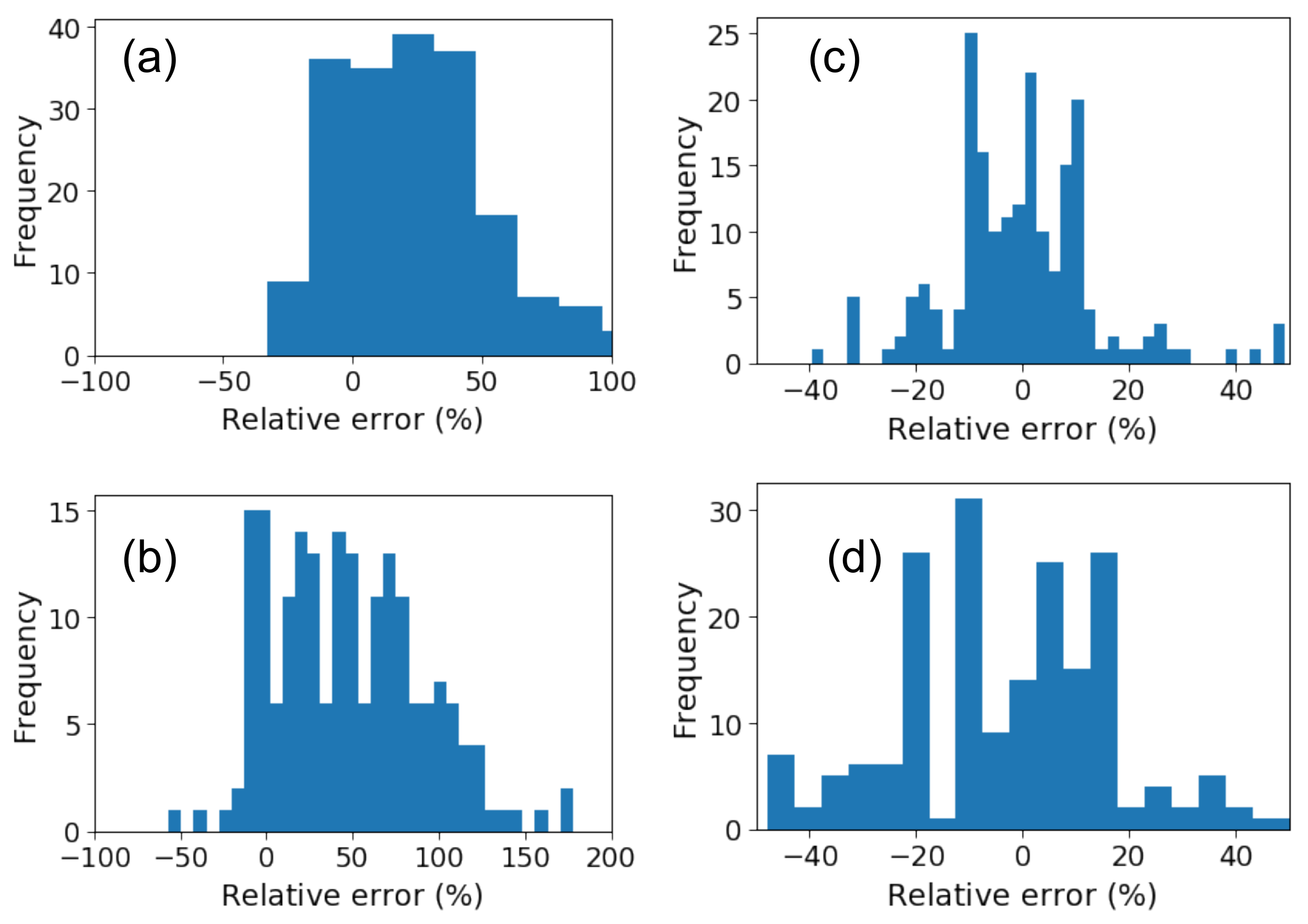}
\caption{Histograms of relative errors for (a) ultimate stress and (b)
  ultimate strain for generated structures with maximizing diversity
  strategy. Histograms of relative errors for (c) ultimate stress and
  (d) ultimate strain for generated structures with maximizing strain
  strategy.  }
\label{fig:figSI10}
\end{figure*}


\bibliography{ml}

\end{document}